\documentclass[aps,pra,reprint,superscriptaddress,floatfix,nofootinbib]{revtex4-1}

\usepackage[english]{babel}
\usepackage[utf8]{inputenc}
\usepackage[T1]{fontenc}

\usepackage{physics}
\usepackage{amsfonts,amsmath,amssymb,accents,braket,mathtools}
\usepackage{dsfont,amsfonts,bm}
\usepackage{array}
\usepackage{enumitem}

\usepackage{placeins}
\usepackage{graphicx}
\graphicspath{ {figures/phases/},{figures/double_well_explanation/},{figures/entanglement/},{figures/number_states/},{figures/densities/}}
\usepackage{epstopdf}

\usepackage[colorlinks=true,citecolor=blue,linkcolor=blue,urlcolor=blue]{hyperref}
\usepackage[capitalise]{cleveref}
\usepackage{color,xcolor}
\usepackage{ulem} 

\begin{document}

\title{Phase separation of a Bose-Bose mixture: impact of the trap and particle number imbalance}

\author{Maxim Pyzh}
	\email{mpyzh@physnet.uni-hamburg.de}
	\affiliation{Zentrum f\"ur Optische Quantentechnologien, Universit\"at
Hamburg, Luruper Chaussee 149, 22761 Hamburg, Germany}
\author{Peter Schmelcher}
	\email{pschmelc@physnet.uni-hamburg.de}
	\affiliation{Zentrum f\"ur Optische Quantentechnologien, Universit\"at 
Hamburg, Luruper Chaussee 149, 22761 Hamburg, Germany}
	\affiliation{The Hamburg Centre for Ultrafast Imaging, Universit\"at 
Hamburg, Luruper Chaussee 149, 22761 Hamburg, Germany}

\begin{abstract}
We explore a few-body mixture of two bosonic species 
confined in quasi-one-dimensional parabolic traps 
of different length scales.
The ground state phase diagrams in the three-dimensional
parameter space spanned by the harmonic length scale ratio, 
inter-species coupling strength and particle number ratio
are investigated. 
As a first case study we use
the mean-field ansatz (MF)
to perform a detailed analysis of the separation mechanism. 
It allows us to derive a simple and intuitive rule
predicting which of the immiscible phases
is energetically more favorable
at the miscible-immiscible phase boundary.
We estimate the critical coupling strength 
for the miscible-immiscible transition and
perform a comparison to correlated many-body results obtained by means
of the Multi-Layer Multi-Configuration Time Dependent 
Hartree method for bosonic mixtures (ML-X).
At a critical ratio of the trap frequencies, 
determined solely by the particle number ratio,
the deviations between MF and ML-X are very pronounced 
and can be attributed to a high degree of entanglement between the components.
As a result, we evidence the breakdown of the effective one-body picture.
Additionally, when many-body correlations play a substantial role,
the one-body density is in general not sufficient
for deciding upon the phase at hand 
which we demonstrate exemplarily.
\end{abstract}

\maketitle

\section{Introduction}
\label{sec:intro}

Binary mixtures of ultra-cold gases have been extensively studied over the past years.
They represent a unique platform for the investigation
of complex interacting many-body quantum systems in a well controlled environment.
In  particular, it is experimentally possible 
to shape the geometry of the trap \cite{TailoredTraps2000},
to reduce the dimensionality
of the relevant motion \cite{1Dgases2008,1Dgases2011}, 
to tune the inter-particle interactions \cite{Feshbach2010,CIR1998,CIR2000,CIR2003,CIR2010}
and prepare samples of only a few atoms \cite{fewbody2012,fewbody2019}.
Numerous experiments have been conducted with different
hyperfine states \cite{Myatt1997,Hall1998,Ketterle1999,Inguscio2000,Aspect2001,
Hall2007,Hirano2010,Becker2008,Engels2011,Oberthaler2015,Hirano2016collision,Hirano2016quench,
dropletsCabrera2018,dropletsInguscio2018},
different elements \cite{Inguscio2002,Weidemuller2002,Inguscio2008,Ospelkaus2008,Cornish2011,Nagerl2011,
Grimm2013,Nagerl2014,Cornish2014,Arlt2015,Wang2015a,Proukakis2018,Wang2015b,Minardi2010} 
or different isotopes \cite{Papp2008,Takahashi2011}
to reveal how the interplay between two condensates 
impacts their stationary properties and non-equilibrium dynamics.
Highlights of these explorations include among others 
the phase separation between the components and symmetry-breaking phenomena
\cite{Hall1998,Papp2008,Hirano2010,Wang2015b,Proukakis2018},
the observation of Efimov physics \cite{Minardi2010} 
and creation of deeply bound dipolar molecules \cite{Ospelkaus2008,Nagerl2014,Cornish2014}, 
as well as dark-bright solitary waves \cite{Becker2008,Engels2011} and
quantum droplets \cite{dropletsCabrera2018,dropletsInguscio2018}.

One of the key properties, 
which makes the multi-component systems attractive and their physics very rich, 
is the miscibility, 
which has significant implications for sympathetic cooling \cite{Aspect2001,Weidemuller2002},
coarse graining dynamics \cite{coarseGraining2004,coarseGraining2008,coarseGraining2010,coarseGraining2014}
and vortex formation \cite{vortex2003,vortex2011} to name a few.
In the very early theoretical investigations a very rich
phase space for the ground state of the Bose-Bose mixture
has been identified.
These investigations \cite{TFAShenoy1996,cGPEBigelow1998,phasesChui2003,phasesOhberg1999,phasesTrippenbach2000} 
are based on the one-body densities obtained 
from solving the underlying mean-field equations, 
commonly known as Gross-Pitaevskii equations.
In case of a weak inter-component coupling one finds a miscible phase 
with a high spatial overlap between the components. 
For a sufficiently large repulsive coupling there are three
types of segragated phases with a rather small overlap.
Two of them are core-shell phases with one component 
being symmetrically surrounded by the other component, 
whereas the third is an asymmetrical phase,
where the rotational or parity symmetry 
of the underlying trapping potential is broken.
Neglecting the kinetic energy (Thomas-Fermi approximation) 
a simple separation criterion 
for the miscible-immiscible transition has been derived
\cite{separationRuleTimmermans1998,separationRuleChui1998,separationRuleEsry1997}.
It depends solely 
on the intra-species and inter-species interactions strengths, which are
easily adjustable by Feshbach or confinement induced resonances
\cite{Feshbach2010,CIR1998,CIR2000,CIR2003,CIR2010}.

However, it has been shown
that this separation criterion, while valid in homogeneous systems, 
should be applied with care
in inhomogeneous geometries. 
There, system parameters
such as trap frequency, particle numbers or mass ratio, 
have also an impact on the miscible-immiscible phase boundary 
\cite{brokenRuleTrapKevrekidis2009,brokenRuleTrapHu2012,brokenRuleTrapProukakis2016,
brokenRuleMassBoronat2018,brokenRuleImbalanceZhang2020}.
From the intuitive point of view 
the trap pressure favors miscibility, 
since it costs energy to extend in space. 
Thus, it requires stronger inter-component repulsion 
for the species to separate. However, 
there are still open questions regarding 
the impact of different length scales,
the characterization of boundaries between the immiscible phases 
and what type of separation will occur 
once the critical coupling is reached.

Another relevant topic affecting the critical coupling strength for a transition 
as well as the resulting type of phase
are the inter-species correlations, which generate entanglement
between the components and lead to bunching of particles of the same species.
Although a mean-field treatment is often justified in experimental setups,
a very thorough numerical analysis of 1D few-body systems has revealed
that an asymmetric immiscible phase is one of the two possible configurations
of an entangled many-body state, the other one being the mirror image.
The one body densities of this so-called composite fermionization phase
\cite{phasesZollner2008,phasesHao2008,phasesPolls2014,phasesZinner2015,phasesPyzh2018}
preserve parity symmetry of the underlying trapping potential and 
have a high spatial overlap, which is uncharacteristic for an immiscible phase.
Nevertheless, the components are indeed separated, which is encoded
in the inter-species two-body density matrix. In experiments,
the single-shots do not represent one-body densities 
but are projections on one of the two mutually exclusive configurations.
An averaging procedure would reveal a parity preserving density, unless
the Hamiltonian itself violates that symmetry, 
such as not coinciding trap centers of the one-body potentials.
Apart from composite fermionization, 
there is a whole class of so called spin-chain phases 
with an even higher degree of entanglement \cite{spinchain2014,spinchain2015,spinchain2016}.
When all interactions in the system become nearly-resonant,
many states become quasi-degenerate and particles, being bosons, 
gain fermionic features like the Pauli exclusion principle.

Considering the above, our work addresses three different points.
First, we characterize the phase diagram 
in a three-dimensional parameter space
spanned by the ratio of the harmonic trap lengths, the inter-species coupling strength
and the particle number ratio.
We switch off intra-component interactions 
to reduce the complexity and gain a better understanding
of the separation process.
A very rich phase diagram is revealed admitting two tri-critical points,
where three phases may coexist.
Second, within the framework of a mean-field approximation,
we perform a detailed analysis of the separation mechanism.
Equipped with this knowledge we derive a selection rule 
for phase separation processes
and a simple algorithm to estimate 
the miscible-immiscible phase boundary.
Finally, we investigate the deviations of the mean-field picture
to a many-body approach. 
For this we use the Multi-Layer Multi-Configuration Time-Dependent 
Hartree method for bosonic mixtures \cite{MLX2013,MLX2013b,MLX2017}.
We find that in the vicinity of the high-entanglement regime
the phase diagram is indeed greatly affected. 
The symmetry-broken phase is replaced by the composite fermionization,
while the onset of symmetry-breaking is linked to the degree of entanglement
reaching a certain threshold.
Furthermore, the location of this beyond mean-field regime strongly depends 
on the harmonic length scale ratio and the particle number ratio.
We also find that the one-body density is in general not sufficient
to distinguish between a core-shell phase and the composite fermionization.

This work is organized as follows. 
In Sec.\ \ref{sec:general_setup} we introduce our physical setup
and in Sec.\ \ref{sec:methodology} our computational approach.
Sec.\ \ref{sec:few_body} is dedicated to a detailed study
of a few-body mixture. 
Sec.\ \ref{sec:few_body_MF} provides intuitive insights 
in the framework of the mean-field approximation,
while Sec.\ \ref{sec:few_body_MLX} focuses on correlation 
and entanglement effects using Multi-Layer Multi-Configuration Time-Dependent 
Hartree method for bosonic mixtures.
The few-to-many-body transition 
is subject of Sec.\ \ref{sec:many_body}.
Finally, we summarize our findings in Sec.\ \ref{sec:conclusions}.

\section{General Framework}
\label{sec:general_setup}

Our system consists of a particle-imbalanced mixture 
of two distinguishable bosonic components, 
denoted by $\sigma \in \{M,I\}$, 
with $N_M$ particles in the majority component and $N_I$ impurities. 
All particles are assumed to be of equal mass $m$
and the intra-component interactions 
are assumed to be zero or negligibly small.
The latter can be achieved by means of
magnetic Feshbach or confinement induced resonances 
\cite{CIR1998,CIR2000,CIR2003,CIR2010,Feshbach2010}.
The majority species interacts with the impurities 
via s-wave contact interaction of coupling strength $g_{_{MI}}$. 
The species are confined in separate quasi-1D harmonic traps 
of different length scales $a_{\sigma}=\sqrt{\hbar/m \omega_{\sigma}}$ 
with trap frequency $\omega_{\sigma}$ and coinciding trap centers.
This can be realized by using two different hyperfine states of $^{87}\rm{Rb}$ 
and species-dependent optical potentials \cite{speciestrap2007,speciestrap2019}. 
We remark that many of the present results can be translated to the mass-imbalanced case.
By choosing $a_{M}$ and $\hbar \omega_{M}$ as length and energy scales 
we arrive at the rescaled Hamiltonian:
\begin{eqnarray}
\label{eq:hamiltonian}
	H && = H_M + H_I + H_{MI} \\
      && = \sum_{i=1}^{N_M} \left( -\frac{1}{2} \frac{\partial^2}{\partial x_i^2} +
			\frac{1}{2} x_i^2 \right)
	     + \sum_{i=1}^{N_I} \left( -\frac{1}{2} \frac{\partial^2}{\partial y_i^2} +
			\frac{1}{2} \eta^2 y_i^2 \right) + \nonumber \\
	  && \quad + g_{_{MI}} \sum_{i=1}^{N_M} \sum_{j=1}^{N_I} \delta(x_i-y_j), \nonumber
\end{eqnarray}
where $x_i$ labels the spatial coordinate of the $i$-th majority particle,
$y_i$ of the $i$-th impurity particle 
and $\eta=\omega_I/\omega_M$ denotes the trap frequency ratio.


In the present work we focus on the ground state characterization
and consider both attractive and repulsive interactions 
ranging from weak to intermediate couplings $g_{_{MI}}\in[-2,2]$
with the impurity being localized 
or delocalized w.r.t.\ the majority species, i.e.\ 
$a_I/a_M=\sqrt{1/\eta} \in [0.5, 1.5]$.
We also study the impact of the particle number ratio $N_I/N_M$
on the system's properties.

\section{Computational Approach}
\label{sec:methodology}

To find the ground state of our binary mixture
we employ imaginary time propagation by means of the
Multi-Layer Multi-Configurational 
Time-Dependent Hartree Method for atomic mixtures (ML-MCTDHX).
For reasons of brevity we call it ML-X from now on.
This multi-configurational wave function based method
for efficiently solving the time-dependent Schroedinger equation 
was first developed for distinguishable degrees of freedom \cite{MCTDH2000}
and ML-X is an extension to indistinguishable particles 
such as bosons or fermions and mixtures thereof \cite{MLX2013,MLX2013b,MLX2017}.
ML-X is an ab-initio method, 
whose power lies in expanding the wave-function 
in time-dependent basis functions.
Let us demonstrate the underlying ansatz for the system at hand:
\begin{equation}
	\ket{\Psi(t)} =\sum_{i=1}^{S}\sqrt{\lambda_i(t)} 
	\ket{\Psi_i^M(t)} \otimes \ket{\Psi_i^I(t)},
\label{eq:wfn_ansatz_species_layer}
\end{equation}
\begin{equation}
	\ket{\Psi_i^{\sigma}(t)} = \sum_{\vec{n}^{\sigma}|N_\sigma}
	C_{i, \vec{n}^{\sigma}}(t)
	\ket{\vec{n}^{\sigma}(t)}.
\label{eq:wfn_ansatz_particle_layer}
\end{equation}

The time-dependent many-body wave-function $\ket{\Psi(t)}$ has two layers: 
the so-called species layer \eqref{eq:wfn_ansatz_species_layer} 
and the particle layer \eqref{eq:wfn_ansatz_particle_layer}. 
In the first step \eqref{eq:wfn_ansatz_species_layer} 
we separate majority and impurity species 
and assign them to $S \in \mathbb{N}$ corresponding 
species wave-functions $\ket{\Psi_i^{\sigma}(t)}$.
The time-dependent coefficients $\lambda_i(t)$ 
are normalized $\sum_{i=1}^{S} \lambda_i(t) = 1$ and
describe the degree of entanglement between the components.
In case $\exists \ i\in\{1,\dots,S\} : \lambda_i(t) \approx 1$ 
the components are said to be disentangled.
In the second step \eqref{eq:wfn_ansatz_particle_layer} 
each species wave-function $\ket{\Psi_i^{\sigma}(t)}$, 
which depends on $N_{\sigma}$ indistinguishable coordinates, 
is expanded in terms of species-dependent symmetrized product states, 
also known as permanents or number states,
$\ket{\vec{n}^{\sigma}}=\ket{n_1^{\sigma},\ldots, n_{s_{\sigma}}^{\sigma}}$
admitting $s_{\sigma}\in \mathbb{N}$ normalized
single particle functions (SPF) $\ket{\varphi_{j}^{\sigma}(t)}$.
The sum is over all possible configurations $\vec{n}^{\sigma}|N_\sigma$
fulfilling the constraint 
$\sum_{i=1}^{s_{\sigma}} n_i^{\sigma} = N_{\sigma}$.
The time dependence of number states is meant implicitly 
through the time-dependence of the underlying SPFs.
Finally, each SPF is represented 
on a primitive one-dimensional time-independent grid \cite{DVR1985}.

When one applies the Dirac-Frenkel variational principle 
\cite{DiracFrenkel2000} to the above ansatz,
one obtains coupled equations of motion 
for the expansion coefficients $\lambda_i(t)$, $C_{i, \vec{n}^{\sigma}}(t)$ 
and the SPFs $\ket{\varphi_{j}^{\sigma}(t)}$.
This procedure allows to considerably reduce the size of the basis set 
as compared to choosing time-independent SPFs constituiting the number states
on the particle layer \eqref{eq:wfn_ansatz_particle_layer}. 
We note that $S=1 \land s_{\sigma}=1$ 
is equivalent to solving coupled Gross-Pitaevskii equations.
We will show parameter regions, where the mean-field description is valid
and regions where it fails as a result of increasing interspecies correlations. 
These generate entanglement between the components and
decrease the degree of condensation of the non-interacting majority atoms.

In the following we will often refer to the 
one-body density $\rho_1^{\sigma}(z)$ of species $\sigma$, 
two-body density matrix $\rho_2^{\sigma}(z,z')$ of species $\sigma$ and
inter-species two-body density matrix $\rho_2^{MI}(x,y)$ of the many-body density operator $\hat{\rho} = \ket{\Psi}\bra{\Psi}$ defined as:
\begin{eqnarray}
	\rho_1^{\sigma}(z) &&= \bra{z} \tr_{_{N_{\sigma} \setminus 1}}
	\{\tr_{_{N_{\bar{\sigma}}}} \{{\hat{\rho}}\}\} \ket{z} \label{eq:rho1}\\
	\rho_2^{\sigma}(z,z') &&= \bra{z,z'} \tr_{_{N_{\sigma} \setminus 2}}
	\{\tr_{_{N_{\bar{\sigma}}}} \{\hat{\rho}\}\} \ket{z,z'} \label{eq:rho2}\\
	\rho_2^{MI}(x,y) &&= \bra{x,y} \tr_{_{N_M \setminus 1}}\{\tr_{_{N_I \setminus 1}} 
	\{\hat{\rho}\}\} \ket{x,y},\label{eq:rho2inter}
\end{eqnarray}
where $N_{\sigma} \setminus n$ stands for integrating out 
$N_{\sigma}-n$ coordinates of component $\sigma$ and $\bar{\sigma} \neq \sigma$.

\section{Phase separation: few body mixture}
\label{sec:few_body}

We start our analysis with a few-body system 
consisting of $N_M = 5$ majority particles with $N_I\in\{1,2\}$ impurities.
First, we uncover the mechanism responsible for the phase separation 
on the mean-field level and subsequently we perform a comparison
to the correlated many-body treatment by means of ML-X.

\subsection{Mean-field approach: Basic mechanism of phase separation}
\label{sec:few_body_MF}

For the mean field description we choose a single species orbital $S=1$, 
yielding a non-entangled state 
$\ket{\Psi(t)} = \ket{\Psi^M(t)} \otimes \ket{\Psi^I(t)}$ 
on the species layer. On the particle layer 
a single SPF $s_{\sigma}=1$ is used for each component,
meaning that particles of the same species are forced to condense 
into the same single-particle state $\varphi^{\sigma}(z,t)$
and $\ket{N_{\sigma}}$ is the only possible number state on the particle level.
Thus, our ansatz is $\ket{\Psi(t)} = \ket{N_{M}(t)} \otimes \ket{N_{I}(t)}$
and only $\varphi^{\sigma}(z,t)$ are time-dependent.
As a result of imaginary time propagation 
we end up with the ground state orbitals $\varphi_{_{MF}}^{\sigma}(z)$.
The interpretation of the mean-field treatment is that each species feels 
in addition to its own external potential 
an averaged one-body potential induced by the other component.
To obtain the effective mean-field Hamiltonian $H_{\sigma}^{MF}$ 
of species $\sigma$ we need to integrate out the other component $\bar{\sigma}$.
For convenience we also subtract the energy offset 
$c_{\bar{\sigma}}=\braket{N_{\bar{\sigma}} | H_{\bar{\sigma}} | N_{\bar{\sigma}}}$ caused by the one-body energy of component $\bar{\sigma}$:
\begin{eqnarray}
\label{eq:MF_hamiltonian}
	\nonumber
	H_{\sigma}^{MF} && = \braket{N_{\bar{\sigma}}|H|N_{\bar{\sigma}}} -
	c_{\bar{\sigma}}
	= H_{\sigma} + N_{\bar{\sigma}} g_{_{MI}} 
	\sum_{i=1}^{N_{\sigma}} \rho_{_{MF}}^{\bar{\sigma}}(z_i) \\
	\nonumber
	&& = \sum_{i=1}^{N_{\sigma}} \left(-\frac{1}{2} \frac{\partial^2}{\partial z_i^2}
	+ V_{\sigma} (z_i) + V_{\sigma}^{ind} (z_i)\right) \\
	&& = \sum_{i=1}^{N_{\sigma}} \left(-\frac{1}{2} \frac{\partial^2}{\partial z_i^2}
	+ V_{\sigma}^{eff} (z_i)\right),
\end{eqnarray}
where $\rho_{_{MF}}^{\sigma}(z) = |\varphi_{_{MF}}^{\sigma}(z)|^2$ 
is the one-body density of species $\sigma$ normalized as 
$\int dz \ \rho_{_{MF}}^{\sigma}(z) = 1$, 
$V_{\sigma}^{ind} (z)=N_{\bar{\sigma}} g_{_{MI}} \rho_{_{MF}}^{\bar{\sigma}}(z)$ 
the induced one-body potential,
$V_{\sigma}^{eff} (z)=V_{\sigma} (z)+V_{\sigma}^{ind} (z)$
the effective one-body potential and $\bar{\sigma}\neq\sigma$.

To systematically distinguish between different phases
we define the following two functions, 
applicable also in the more general case of a many-body treatment 
in Sec.\ \ref{sec:few_body_MLX}:
\begin{equation}
	\label{quali1}
	\Delta_{\sigma} = 
	\frac{\rho^{\sigma}_{1} (z=0)}{\underset{z}{\rm{max}} \ \rho^{\sigma}_1 (z)}
\end{equation}
\begin{equation}
	\label{quali2}
	d = |\int_{-\infty}^{\infty} dz \ z \rho_{1}^{M}(z) - 
	\int_{-\infty}^{\infty} dz \ z \rho_{1}^{I}(z)|,
\end{equation}
with the one-body density $\rho_1^{\sigma}(z)$ 
of component $\sigma$ \eqref{eq:rho1}.
Eq.\ \eqref{quali1} compares the one-body density $\rho_1^{\sigma}(z)$ 
at the trap center with its maximum value, 
while eq.\ \eqref{quali2} checks for parity asymmetry, 
as we will argue below.
The above equations are motivated from the literature 
on binary mixtures and we provide a brief summary
on the discovered ground state phases and some of their properties,
which will be relevant in the following discussions.

For weak couplings there is a miscible phase $M$ 
with a high spatial overlap 
of the one-body densities $\rho_1^{\sigma}(z)$. 
As a result both components exhibit a Gaussian profile ($\Delta_{\sigma}=1$)
and occupy the center of their trap ($d=0$). 
The state is disentangled and both species are condensed.
For negative couplings, i.e.\ attractive interactions,
the phase remains miscible
and the widths of the Gaussian densities shrink with decreasing coupling strength.
For stronger positive couplings 
three different phase separation scenarios are possible.
In case the majority species occupies the trap center ($\Delta_{M}=1$),
pushing the impurities outside 
in a way that the impurity density forms a shell 
around the majority density
with two parity-symmetric humps ($\Delta_{I}<1$ and $d=0$),
we have a core-shell $IMI$ phase.
When the impurities remain at the trap center instead ($\Delta_{I}=1$) 
with the majority species forming a shell ($\Delta_{M}<1$ and $d=0$), 
we have a core-shell $MIM$ phase.
Finally, when both species develop 
two parity-symmetric humps
with a local minimum at the trap center ($\Delta_{\sigma}<1$ and $d=0$)
we have a composite fermionization phase $CF$.
On the level of one-body densities 
$CF$ appears to be miscible owing to the high spatial overlap
between the components.
However, the deviations to the miscible phase become evident upon investigating
the two-body density matrices \eqref{eq:rho2} and \eqref{eq:rho2inter}.
Namely, two particles of the same component can be found
either on the left or the right side w.r.t.\ trap center,
while two particles of different components are always on opposite sides.

While the core-shell phases $IMI$ and $MIM$ 
do not rely on entanglement between the components, 
$CF$ is always an entangled many-body state
made out of two major species orbitals $S=2$ 
and two major SPFs $s_{\sigma}=2$ on the particle layer.
Thus, $CF$ cannot be obtained within the mean-field approximation.
In fact, we observe that once the entanglement 
of the true many-body state, 
characterized by the von Neumann entropy 
(see Sec.\ \ref{sec:few_body_MLX}), 
reaches a certain threshold, 
a collapse to a phase with broken parity symmetry 
($d>0$) will take place
in the mean-field picture. 
We abbreviate this phase with $SB$ from now on.

The origin of $SB$
is the onset of a quasi-degeneracy between the ground state 
and the first excited state
of the many-body spectrum, which becomes an exact degeneracy 
in the limit of $g_{_{MI}} \to \infty$. 
Once this limit is reached, 
any superposition of those two states 
is also an eigenstate of \eqref{eq:hamiltonian}. 
Since they are of different parity symmetry $P$ and $[H,P]=0$, 
it is possible to choose the superposition 
to be parity symmetric or to break 
the parity symmetry of \eqref{eq:hamiltonian}.
It was suggested \cite{phasesZollner2008} that
the corresponding many-body wave-function 
may be written in terms of number states as
$\ket{\Psi} =c_1 \ket{N_M}_L \ket{0_M}_R \otimes \ket{0_I}_L \ket{N_I}_R
+ c_2 \ket{0_M}_L \ket{N_M}_R \otimes \ket{N_I}_L \ket{0_I}_R$
with two parity-broken SPFs $\varphi^{\sigma}_j(z)$
featuring an asymmetric Gaussian shape 
with a maximum on the j=L(eft) or j=R(ight) side w.r.t.\ the trap center.
Within the mean-field approximation
the eigenenergy of the first excited state 
coincides with the ground state energy
already for a finite coupling $g_{_{MI}}$.
Since mean-field does not incorporate entanglement, 
the state collapses either to
$\ket{N_M}_L \otimes \ket{N_I}_R$ 
or to $\ket{N_M}_R \otimes \ket{N_I}_L$,
resulting in a phase with broken parity symmetry.

With this we end our overview over different phases and showcase 
a compact summary of the phases:
\begin{equation} 
\label{eq:classification}
	\begin{cases} 
	\text{M} & \mbox{:} \
	d = 0 \ \land \Delta_{_{M}} = 1 \ \land \Delta_{_{I}} = 1 \\
	\text{IMI} & \mbox{:} \
	d = 0 \ \land \Delta_{_{M}} = 1 \ \land \Delta_{_{I}} < 1 \\
	\text{MIM} & \mbox{:} \
	d = 0 \ \land \Delta_{_{M}} < 1 \ \land \Delta_{_{I}} = 1 \\
	\text{CF} & \mbox{:} \
	d = 0 \ \land \Delta_{_{M}} < 1 \ \land \Delta_{_{I}} < 1 \\
	\text{SB} &\mbox{:} \
	d > 0 \ \land \Delta_{_{M}} < 1 \ \land \Delta_{_{I}} < 1
	\end{cases}
\end{equation}

\begin{figure*}[ht]
	\centering
	\includegraphics[width=0.75\linewidth]{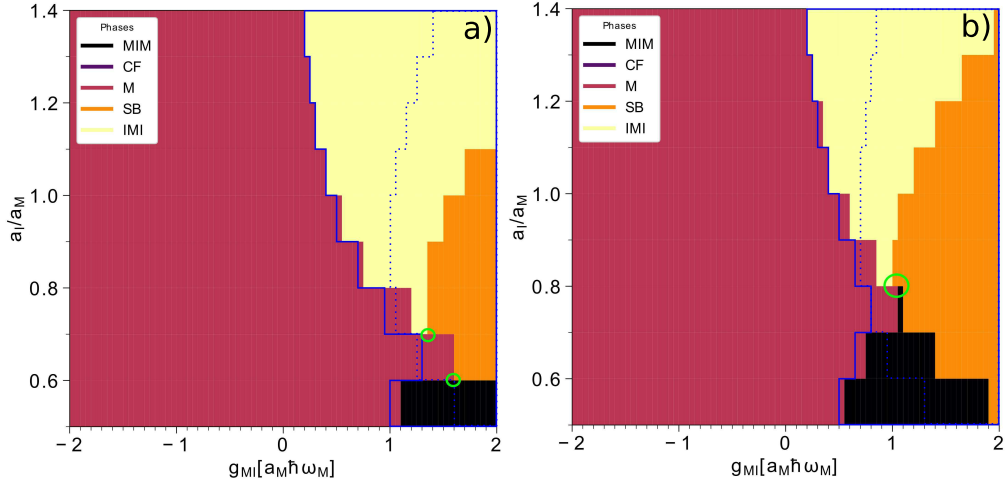}
	\caption{Mean-field ground state phase diagram 
	for $N_B=5$ majority particles 
	and a) $N_I=1$ 
	or b) $N_I=2$ impurities
	as a function of the inter-component coupling strength $g_{_{MI}}$
	and impurity localization $a_{I}/a_{M}=\sqrt{1/\eta}$ 
	with $\eta=\omega_{I}/\omega_{M}$ being the trap frequency ratio 
	and $a_{\sigma}=\sqrt{\hbar/m \omega_{\sigma}}$ 
	the harmonic oscillator length of species $\sigma$. 
	The nomenclature of phases is as follows: 
	$M$ for miscible, 
	$MIM$ for core-shell with impurity at the core, 
	$IMI$ for core-shell with majority at the core,
	$CF$ for composite fermionization and
	$SB$ for a phase with broken parity symmetry.
	The blue solid curve represents
	the miscible-immiscible phase boundary 
	according to \eqref{eq:critical_g}. The blue dotted line
	is an estimate for the $SB$ phase boundary
	according to \eqref{eq:SP_boundary}. 
	Green circles indicate tri-critical points.
	The coarse structure is due to the finite step-size of our data w.r.t.\ $a_I/a_M$.}
	\label{fig:MF_few_body.eps}
\end{figure*}

In Fig.\ \ref{fig:MF_few_body.eps} 
we depict the ground state phases 
within the mean-field approximation for $N_B=5$ 
with a) $N_I=1$ and b) $N_I=2$ impurities
as a function of the inter-component coupling strength $g_{_{MI}}$
and the impurity localization $a_{I}/a_{M}$. 
As expected $CF$ is not among the phases 
in Fig.\ \ref{fig:MF_few_body.eps}.
The transition region on the $a_{I}/a_{M}$ axis,
where core-shell $MIM$ 
is replaced by core-shell $IMI$,
can be tuned by variation of the particle number ratio such that
for $N_I=N_M$ it lies at $a_{I}/a_{M}=1$ (not shown), 
while for increasing particle imbalance  $N_I/N_M<1$
it is shifted towards a lower $a_{I}/a_{M}$ ratio.
This is also the point, where the coupling strength $g_{_{MI}}$,
required for the realization of the $SB$ phase, is the smallest.
We will see later in Sec.\ \ref{sec:few_body_MLX} that
the species entropy has here its global maximum. 
Note that each phase diagram features
critical points, where three different phases can coexist 
(green circles).

Now that we have identified the phases,
we are going to shed some light on the mechanism behind 
the phase separation 
taking place for different specific coupling strength $g_{_{MI}}$
for a fixed trap ratio $\eta$.
In particular, we will provide a simple formula, 
which determines which of the 
core-shell structures is energetically more favorable.
Additionally, we provide an estimate on the miscible-immiscible 
transition region and on the $SB$ phase boundary.

\begin{figure*}[ht]
	\centering
	\includegraphics[width=1.0\linewidth]{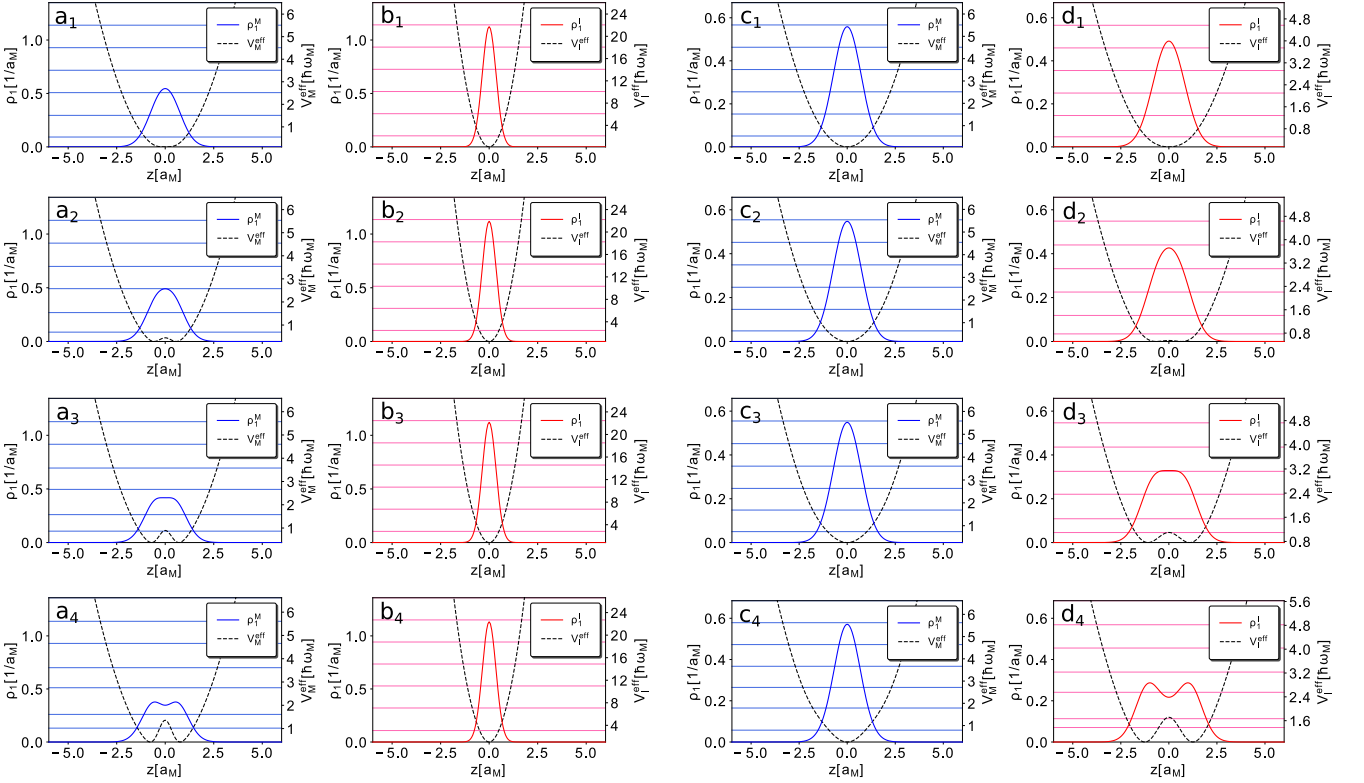}
	\caption{Ground state densities $\rho_{_{MF}}^{\sigma}(z)$ inside
	the induced one-body potential $V_{\sigma}^{eff}(z)$ 
	from \eqref{eq:MF_hamiltonian} for
	$N_M=5$, $N_I=2$ and either 
    $a_I/a_M=0.5$ (columns 1 and 2 for majority 
    and impurity species respectively) or
    $a_I/a_M=1.1$ (columns 3 and 4 for majority and impurity 
    species respectively).
	Rows from top to bottom 
	correspond to a variation of the inter-component coupling 
	$g_{_{MI}}\in\{0.05,0.2,0.4,0.6\}$.
	Horizontal lines depict eigenenergies of \eqref{eq:MF_hamiltonian}.}
	\label{fig:dw_picture.eps}
\end{figure*}

Let us make two horizontal cuts
across the phase diagram of Fig.\ \ref{fig:MF_few_body.eps}b)
at $a_I/a_M=0.5$ and at $a_I/a_M=1.1$. 
In Fig.\ \ref{fig:dw_picture.eps} we take a closer look at the variation 
of the one-body densities $\rho_{_{MF}}^{\sigma}$
being part the effective one-body potential 
$V_{\sigma}^{eff}$ \eqref{eq:MF_hamiltonian}
when increasing the coupling strength $g_{_{MI}}$.
First, let us focus on columns 1 and 2, corresponding to $a_I/a_M=0.5$.
For very weak coupling (first row), 
every atom to a good approximation populates the energetically lowest
harmonic oscillator orbital of the respective potential $V_{\sigma}$.
The induced potential $V_{\sigma}^{ind}$ 
gains an amplitude linearly with $g_{_{MI}}$
and with the density profiles being Gaussians of different widths
we observe the appearance of a small barrier 
in $V_M^{eff}$ at $g_{_{MI}}=0.2$ 
(Fig.\ \ref{fig:dw_picture.eps} a2).
This barrier grows with $g_{_{MI}}$ and at $g_{_{MI}}=0.4$ 
(Fig.\ \ref{fig:dw_picture.eps} a3) 
it becomes comparable to the ground state energy 
of the effective potential,
while the one-body density $\rho_{_{MF}}^M$ 
turns flat at the trap origin. 
Once the ground state energy drops below the barrier height, 
two density humps appear and core-shell $MIM$ is established
(Fig.\ \ref{fig:dw_picture.eps} a4).
Meanwhile, the effective potential of the impurity $V_I^{eff}$
does not show significant deviations from the harmonic case 
(second column).
Especially, the induced part $V_I^{ind}$, 
being initially also a Gaussian,
is not capable to produce a barrier at the trap center.
Similar statements can be made for columns 3 and 4, 
corresponding to $a_I/a_M=1.1$.
The only difference is that $V_I^{eff}$ 
develops a barrier instead, whereas
$V_M^{eff}$ shows only a slight variation, 
which finally leads to the core-shell $IMI$ phase.

Motivated by the above observation we define 
an alternative phase classification from an energetical point of view:
\begin{equation}
\label{eq:classification_energy}
	\begin{cases} 
	\text{M} & \mbox{:} \
	E_{_{0,\sigma}}^{MF}-V_{\sigma}^{eff}(0) > 0 \\
	\text{IMI} & \mbox{:} \
	E_{_{0,M}}^{MF}-V_{M}^{eff}(0) > 0 
    \ \land \
	E_{_{0,I}}^{MF}-V_{I}^{eff}(0) < 0  \\
	\text{MIM} & \mbox{:} \
	E_{_{0,M}}^{MF}-V_{M}^{eff}(0) < 0 
    \ \land \
	E_{_{0,I}}^{MF}-V_{I}^{eff}(0) > 0  \\
	\text{SB} &\mbox{:} \
	d > 0
	\end{cases}
\end{equation}
where $E_{_{0,\sigma}}^{MF}$ is the ground state energy of \eqref{eq:MF_hamiltonian}. 
As long as the ground state energy
of the effective species Hamiltonian exceeds the effective potential height 
at the trap center, the species remains at the trap center.
Phase diagrams produced this way
match exactly the ones shown in Fig.\ \ref{fig:MF_few_body.eps}.

The interpretation is now as follows. 
For a very weak coupling both the majority and the impurity 
reside in the ground state of the harmonic oscillator.
Once the induced potential $V_{\sigma}^{ind}$ of species $\sigma$ 
becomes large enough to produce a barrier in $V_{\sigma}^{eff}$, 
the corresponding density $\rho_{_{MF}}^{\sigma}$ will start to expand. 
By growing in width it will prevent the other component $\bar{\sigma}$
from developing a barrier of its own. 
When the height of the potential barrier becomes of the same magnitude 
as the lowest energy
of the corresponding effective potential, 
the species $\sigma$ splits into two fragments.  
Then it starts squeezing the other component $\bar{\sigma}$
by increasing the effective trap frequency 
of the renormalized harmonic oscillator $V_{\bar{\sigma}}^{eff}$.

The barrier in $V_{\sigma}^{eff}$ appears 
once the following condition is fulfilled:
\begin{equation}
\label{eq:barrier_criterion}
	\exists \ x_0 \neq 0 : \frac{d}{dx} V_{\sigma}^{eff} \Bigr\rvert_{x_0}= 0.
\end{equation}
Assuming one-body densities to be unperturbed harmonic oscillator ground states, we obtain the following effective potentials:
\begin{eqnarray}
\label{eq:barrier_criterion_M_pot}
	&& V_{M}^{eff}(z) \approx \frac{1}{2}z^2 
	+ g_{_{MI}} N_I \sqrt{\frac{\eta}{\pi}} e^{-\eta z^2} \\
\label{eq:barrier_criterion_I_pot}
	&& V_{I}^{eff}(z) \approx \frac{1}{2}\eta^2 z^2 
	+ g_{_{MI}} N_M \sqrt{\frac{1}{\pi}} e^{-z^2},
\end{eqnarray}
and the corresponding barrier conditions:
\begin{eqnarray}
\label{eq:barrier_criterion_M}
	&& \frac{\sqrt{\pi}}{2 N_I \sqrt{\eta^3}} 	
	\hat{=} g_{_{MI}}^M < g_{_{MI}} \\
\label{eq:barrier_criterion_I}
	&&  \frac{\sqrt{\pi} \eta^2}{2 N_M} 
	\hat{=} g_{_{MI}}^I< g_{_{MI}}.
\end{eqnarray}
For given particle numbers $N_M$, $N_I$ and trap ratio $\eta$ 
either condition \eqref{eq:barrier_criterion_M} 
or condition \eqref{eq:barrier_criterion_I} 
will be fulfilled first upon increasing the coupling $g_{_{MI}}$
and thus either the majority or the impurity will form a shell.
We remark that the above criterion for barrier formation
is inversely proportional to the particle number of the other component, 
while the dependence on the trap ratio $\eta$ for the majority 
differs substantially from the one for the impurity.
Furthermore, for a fixed particle number ratio 
there is a critical trap ratio $\eta_c$, for which 
\eqref{eq:barrier_criterion_M} and \eqref{eq:barrier_criterion_I}
can be fulfilled simultaneously.
\begin{equation}
	\sqrt{1/\eta_c}=\sqrt[7]{N_I/N_M}
\end{equation}
Around this critical region we expect that none of the components
will occupy the trap center. 
We summarize our findings in a simple formula, which determines the
type of phase separation at the miscible-immiscible phase boundary:
\begin{equation}
\label{eq:critical_trap}
	\begin{cases} 
	\text{core shell MIM} & \mbox{:} \
	\eta \gg \eta_c \\
	\text{core shell IMI} & \mbox{:} \
	\eta \ll \eta_c  \\
	\text{CF or SB} & \mbox{:} \
	\eta \approx \eta_c \\
	\end{cases}
\end{equation}
For particle number ratios discussed in this section, 
the critical region lies at $a_I/a_M \approx 0.8$ (Fig.\ \ref{fig:MF_few_body.eps}a) 
and at $a_I/a_M \approx 0.9$ (Fig.\ \ref{fig:MF_few_body.eps}b). 

Next, we want to find an estimate for the miscible-immiscible
phase boundary $g^c_{_{MI}}$. To this end we combine 
the energetical separation criterion 
in \eqref{eq:classification_energy} 
with approximate effective potentials from 
\eqref{eq:barrier_criterion_M_pot} and \eqref{eq:barrier_criterion_I_pot}.
Specifically, for a given particle number ratio $N_I/N_M$
we determine the critical trap ratio $\eta_c$.
Then depending on the choice of $\eta$ we
solve numerically for the ground state energy 
of a single particle inside the effective potential
\eqref{eq:barrier_criterion_M_pot} or \eqref{eq:barrier_criterion_I_pot}.
Finally, we compare this energy to the potential height at the trap center:
\begin{widetext}
\begin{equation}
\label{eq:critical_g}
	\begin{cases} 
	a_I/a_M < \sqrt[7]{N_I/N_M} & \mbox{:} 
	\ H_M^{eff}= -\frac{1}{2} \frac{\partial^2}{\partial x^2} + V_{M}^{eff}(x)
		\begin{cases}
		E^{eff}_{_{0,M}} > g_{_{MI}} N_I \sqrt{\frac{\eta}{\pi}}  
		& \Rightarrow \ \text{M} \\
		E^{eff}_{_{0,M}} < g_{_{MI}} N_I \sqrt{\frac{\eta}{\pi}}  
		& \Rightarrow \ \text{MIM} \\
		\end{cases}	\\
	\\
	a_I/a_M > \sqrt[7]{N_I/N_M} & \mbox{:} 
	\ H_I^{eff}=-\frac{1}{2} \frac{\partial^2}{\partial y^2} + V_{I}^{eff}(y)
		\begin{cases}
		E^{eff}_{_{0,I}} > g_{_{MI}} N_B \sqrt{\frac{1}{\pi}}  
		& \Rightarrow \ \text{M} \\
		E^{eff}_{_{0,I}} < g_{_{MI}} N_B \sqrt{\frac{1}{\pi}}  
		& \Rightarrow \ \text{IMI} \\
		\end{cases}	\\
	\end{cases}
\end{equation}
\end{widetext}
The results are plotted as blue solid curves 
in Fig. \ref{fig:MF_few_body.eps}.
We recognize that it performs quite well 
except for $\eta \approx \eta_c$,
where it underestimates $g^c_{_{MI}}$.

We can also get a rough estimate on the $SB$ phase boundary 
$g_{_{MI}}^{^{SB}}$ by
using the following Gaussian ansatz: 
\begin{equation}
	\varphi^{\sigma}(z)= \sqrt[4]{\frac{\beta_{\sigma}}{\pi}} 
	e^{-\frac{\beta_{\sigma}(z-z_\sigma)^2}{2}},
\end{equation}
with the width $\beta_{\sigma}$ and the displacement $z_\sigma$ 
of component $\sigma$
being variational parameters.
We evaluate the expectation value of \eqref{eq:hamiltonian} and 
minimize the energy w.r.t.\ the above variation parameters.
By looking further at the special case 
when the relative position $|z_M-z_I|$
becomes zero one arrives after some algebraic transformations at:
\begin{equation}
	\label{eq:SP_boundary}
	g_{_{MI}}^{^{SB}} N_I = \frac{\sqrt{\pi}}{2 \eta} 
	\sqrt[4]{\frac{\gamma}{1+\gamma \eta^2}} 
	(1+\eta^2 \sqrt{\gamma})^{\frac{3}{2}},
\end{equation}
with particle number ratio $\gamma=N_I/N_M$. 
We remark that this equation
reduces to eq.\ (8) from \cite{brokenRuleImbalanceZhang2020} for $\eta=1$.
Although this equation describes well 
the qualitative behavior of the $SB$ phase boundary,
quantitatively it scales badly 
when the trap ratio $\eta$ deviates from $\eta_c$ (blue dotted line in Fig.\ \ref{fig:MF_few_body.eps}). 
There are two possible reasons for this. 
First, our ansatz incorporates only $M$ and $SB$ phases, 
while ignoring the core-shell phases. 
Thus, as one draws away from $\eta_c$ 
the core-shell parameter region, 
which lies in-between $M$ and
$SB$, grows in size
making the estimate inefficient.
The other reason is
that the mean-field solution $\varphi^{\sigma}_{_{MF}}$
of the $SB$ phase is rather an asymmetric Gaussian.

Finally, we discuss the limiting cases. 
When $\eta \rightarrow \infty$ ($a_{I}/a_{M} \rightarrow 0$)
the impurity becomes highly localized at $z=0$. 
It will not be affected by the majority atoms.
Meanwhile the majority species will be subject to an additional delta-potential
at $z=0$ with potential strength $g_{_{MI}} N_I$. 
This analytically solvable one-body problem 
results in a Weber differential equation.
Upon increasing the delta-potential pre-factor $g_{_{MI}} N_I$ 
the initially unperturbed Gaussian solution develops 
a cusp at the trap center, 
whose depth tends to zero as the pre-factor goes to infinity.
When $\eta \rightarrow 0$ ($a_{I}/a_{M} \rightarrow \infty$),
we can change our perspective 
by rescaling the Hamiltonian in impurity harmonic units
and argue in a similar way as above.

In the following section we compare to the results obtained for the
corresponding correlated many-body approach of ML-X.

\subsection{ML-X: modifications of the phase diagram due to correlations and entanglement}
\label{sec:few_body_MLX}

\begin{figure*}[ht]
	\centering
	\includegraphics[width=0.75\linewidth]{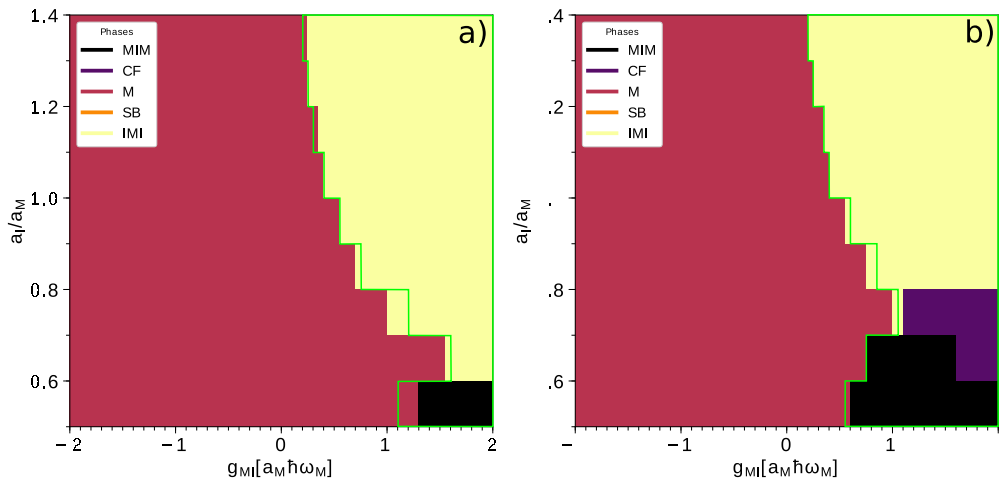}
	\caption{ML-X ground state phase diagram 
	for $N_B=5$ majority particles 
	and a) $N_I=1$ or b) $N_I=2$ impurities
	as a function of the inter-component coupling strength $g_{_{MI}}$
	and impurity localization $a_{I}/a_{M}=\sqrt{1/\eta}$ 
	with $\eta=\omega_{I}/\omega_{M}$ being the trap frequency ratio 
	and $a_{\sigma}=\sqrt{\hbar/m \omega_{\sigma}}$ 
	the harmonic oscillator length of species $\sigma$.
	The nomenclature of phases is as follows: 
	$M$ for miscible, 
	$MIM$ for core-shell with impurity at the core, 
	$IMI$ for core-shell with majority at the core,
	$CF$ for composite fermionization and
	$SB$ for a phase with broken parity symmetry.
	The green solid curve 
	represents the miscible-immiscible phase boundary 
	based on the mean-field treatment.
	The coarse structure is due to the finite step-size of our data w.r.t.\ $a_I/a_M$.}
	\label{fig:MLX_few_body.eps}
\end{figure*}

For the total wave-function in \eqref{eq:wfn_ansatz_species_layer}
we use $S=8$ species orbitals 
and $s_{\sigma}=8$ SPFs for each component.
We perform again an imaginary time-propagation of
an initially chosen wave-function
and obtain the ground state of \eqref{eq:hamiltonian}.
In Fig.\ \ref{fig:MLX_few_body.eps} 
we show the resulting ground state phases
based on the selection rules \eqref{eq:classification} 
for $N_B=5$ and a) $N_I=1$ or b) $N_I=2$. 
We remark that the alternative selection scheme 
defined in \eqref{eq:classification_energy} does not apply here and
below we provide an explanation why it fails.
The first eye-catching feature
is that the $SB$ phase has completely disappeared, 
as expected, 
since it is an artifact of the mean-field treatment.
Additionally, we observe the presence of composite fermionization $CF$ 
for the case of two impurities
in Fig.\ \ref{fig:MLX_few_body.eps} b).
Overall, the transition between the miscible phase 
and separated phases takes places
at a different coupling strength $g_{_{MI}}^c$ 
for a fixed trap ratio $\eta$.

\begin{figure*}[ht]
	\centering
	\includegraphics[width=0.75\linewidth]{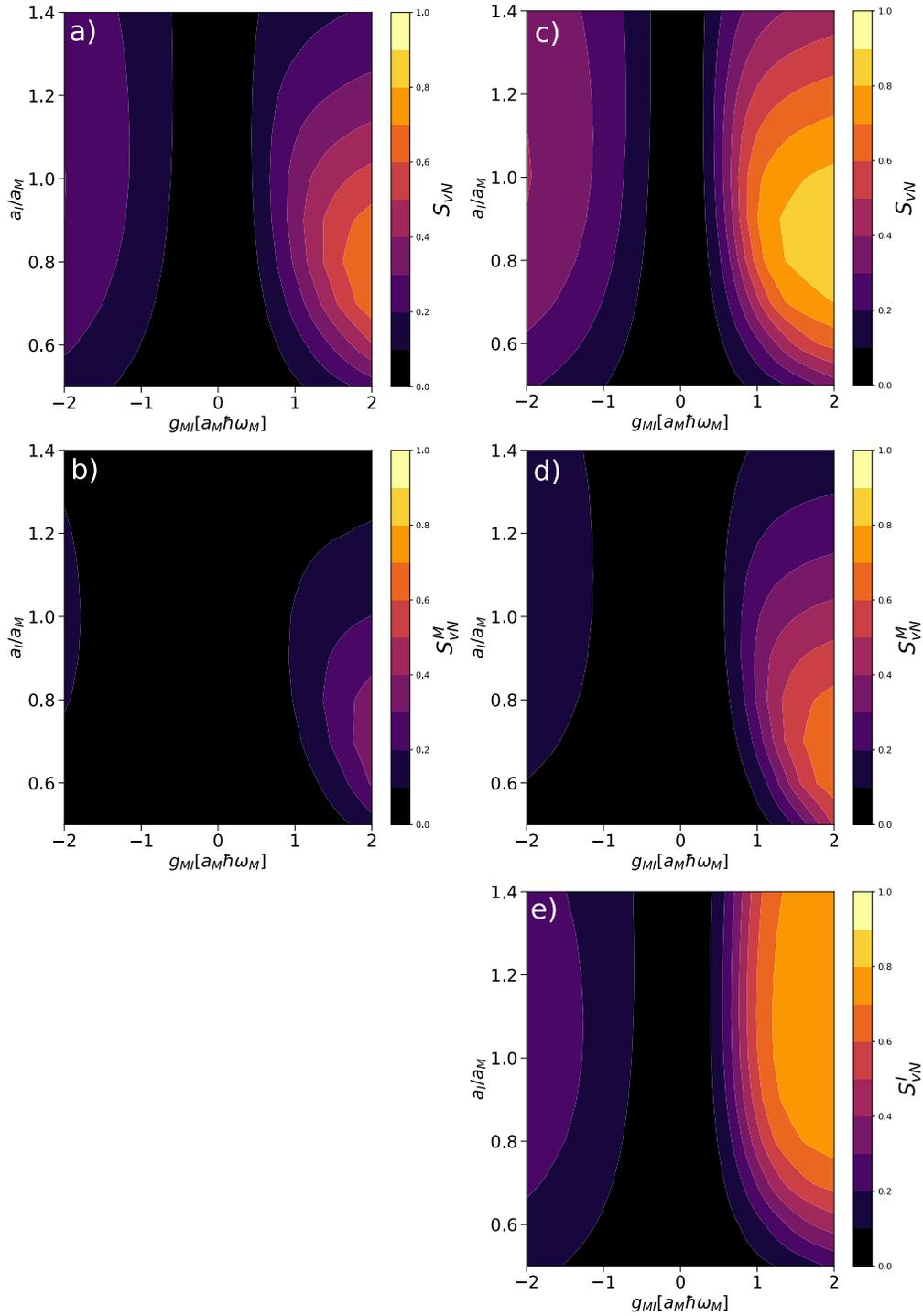}
	\caption{Ground state (species) entanglement entropy $S_{vN}$
	from eq.\ \eqref{eq:species_entropy} (first row) 
	and the fragmentation entropy $S_{vN}^{\sigma}$
	from eq.\ \eqref{eq:particle_entropy} 
	for the majority (second row) and impurity (third row)
	for $N_B=5$ majority particles 
	and $N_I=1$ (first column) or $N_I=2$ (second column) impurities
	as a function of the inter-component coupling strength $g_{_{MI}}$
	and impurity localization $a_{I}/a_{M}=\sqrt{1/\eta}$ 
	with $\eta=\omega_{I}/\omega_{M}$ being the trap frequency ratio 
	and $a_{\sigma}=\sqrt{\hbar/m \omega_{\sigma}}$ 
	the harmonic oscillator length of species $\sigma$.}
	\label{fig:entanglement.eps}
\end{figure*}

In order to better understand why the phase diagram is altered this way,
we investigate in Fig.\ \ref{fig:entanglement.eps} 
the von-Neumann entropy $S_{vN}$ on the species layer (first row)
as well as
the von-Neumann entropy of the majority species $S_{vN}^{M}$ (second row)
and the impurity species $S_{vN}^{I}$ (third row).
$S_{vN}$ characterizes the degree of entanglement between the components
(entanglement entropy)
while $S_{vN}^{\sigma}$ reflects the degree of species fragmentation
(fragmentation entropy).
The definitions are as follows:
\begin{eqnarray}
\label{eq:species_entropy}
	S_{vN} && = - \sum_{i=1}^{S} \lambda_i \ln{\lambda_i} \\
\label{eq:particle_entropy}
	S_{vN}^{\sigma} && = - \sum_{i=1}^{s_{\sigma}} m_i \ln{m_i} \; \;
	\text{with} \ \hat{\rho}_1^{\sigma}
	=\sum_{i=1}^{s_{\sigma}} m_i \ket{m_i}\bra{m_i}
\end{eqnarray}
where $\lambda_i$ are expansion coefficients 
from \eqref{eq:wfn_ansatz_species_layer},
$m_i$ natural populations satisfying $\sum_{i=1}^{s_{\sigma}}m_i=1$
and $\ket{m_i}$ natural orbitals 
of the spectrally decomposed one-body density operator
$\hat{\rho}_1^{\sigma}$.
The entanglement entropy is bounded 
by the equal distribution of orbitals $S_{vN} \leq \ln(S)$,
whereas for two dominantly occupied orbitals
we expect $S_{vN} \leq \ln(2)\approx0.7$.
If $S_{vN}=0$, then there is no entanglement between the species 
and the wave-function
is a simple product state on the species layer. 
Similarly, fragmentation entropy $S_{vN}^{\sigma}=0$
means that all particles occupy the same SPF
and the species is thus condensed.
For parameter values where this is fulfilled
a mean-field treatment is well justified.
However, in Fig.\ \ref{fig:entanglement.eps} we recognize
that for stronger couplings $g_{_{MI}}$ this is not the case. 
Particularly, in the vicinity of the critical region 
$a_I/a_M \approx \sqrt[7]{N_I/N_M}$ at positive $g_{_{MI}}$, 
identified in the previous section
as highly competitive, the entanglement entropy 
$S_{vN}$ is very pronounced
(Fig.\ \ref{fig:entanglement.eps} first row).
The fragmentation entropy of the majority species $S_{vN}^M$
is comparatively weaker and slightly shifted
towards a smaller length scale ratio $a_I/a_M$ at positive $g_{_{MI}}$
(Fig.\ \ref{fig:entanglement.eps} second row).
The fragmentation entropy of the impurity species $S_{vN}^I$
for $N_I=1$ (not shown) coincides with the entanglement entropy $S_{vN}$
(Fig.\ \ref{fig:entanglement.eps}a), while
for $N_I=2$ there are substantial differences 
(see Fig.\ \ref{fig:entanglement.eps}e).
Namely,
the impurity shows a higher degree of fragmentation 
when it is less confined compared to the majority
species and vice versa.
In contrast to positive couplings $g_{_{MI}}$, for negative couplings the entanglement
and species fragmentation build up with a much slower rate.
Finally, we emphasize that phase separation like core-shell
$MIM$ or $IMI$ are not necessarily 
related to a high degree of entanglement or species depletion,
whereas $CF$ is located in the parameter region, 
where $S_{vN}$ takes the highest values.
Another striking observation is that the onset of the $SB$ phase
from Fig.\ \ref{fig:MF_few_body.eps} is related to the
entanglement entropy reaching some threshold value 
around $S_{vN}\approx0.5$  at positive couplings $g_{_{MI}}$
(compare to Fig.\ \ref{fig:entanglement.eps} first row).

Now that we have identified the parameter space 
where deviations from mean-field are to be expected, 
we want to gain a deeper insight into how the effective picture is affected
as a result of increasing correlations. 
For this purpose we define an effective single-body Hamiltonian of species $\sigma$
similar to the one in \eqref{eq:MF_hamiltonian}, 
except that we use the exact many-body densities $\rho_1^{\sigma}$ 
instead of the mean-field densities $\rho_{_{MF}}^{\sigma}$:
\begin{eqnarray}
\label{eq:projection_hamiltonian}
	\nonumber
	H_{\sigma}^{eff} && = H_{\sigma} + N_{\bar{\sigma}} g_{_{MI}} 
	\sum_{i=1}^{N_{\sigma}} \rho_1^{\bar{\sigma}}(z_i) 
	\qquad \text{with} \ \bar{\sigma}\neq\sigma\\
	&&= \sum_{i=1}^{N_{\sigma}} \left(-\frac{1}{2} \frac{\partial^2}{\partial z_i^2}
	+ V_{\sigma}^{eff} (z_i)\right)
\end{eqnarray}
Next, we diagonalize \eqref{eq:projection_hamiltonian} 
and use the obtained eigenfunctions $\tilde{\varphi}_i^{\sigma}$
as SPFs for number states $\ket{\vec{n}^{M}} \otimes \ket{\vec{n}^{I}}$
on which we project our many-body ground state $\ket{\Psi}$.
The reader should distinguish the latter SPFs $\tilde{\varphi}_i^{\sigma}$
from the numerical SPFs  $\varphi_i^{\sigma}$ 
obtained by improved relaxation which define the permanents
contained in our ML-X total wave-function.
Thus, we decompose our ground state in terms of disentangled product states
made out of single permanents.
We anticipate that $\ket{N_M}\ket{N_I}$ 
represents dominant contribution to $\ket{\Psi}$,
which should be the case whenever a mean-field approach is valid.
From the previous analysis we observed 
that the entanglement entropy values
were mostly $S_{vN}^{\sigma} \leq 0.7$, 
which suggests two relevant SPFs. 
Indeed, our many-body state consists of two major orbitals
and two major SPFs.
Furthermore, taking parity symmetry into account 
and considering at most two-particle excitations,
we conclude that number states $\ket{N_M-1,1}\ket{N_I-1,1}$, 
$\ket{N_M-2,2}\ket{N_I}$ and $\ket{N_M}\ket{N_I-2,2}$ 
may become of relevance too at stronger couplings.
We remark that the one-body density operator
of number state $\ket{N_{\sigma}-n_2^{\sigma},n_2^{\sigma}}$
with $n_2^{\sigma}$ particles in the odd orbital $\tilde{\varphi}_2^{\sigma}$
will be a mixed state of one even and one odd orbital,
eventually featuring two humps 
in the corresponding one-body density.
Thus, depending on the occupation amplitude of such states,
they may either accelerate or slow down the development of humps 
in $\rho_1^{\sigma}(z)$
thereby quantitatively shifting the critical coupling $g^c_{_{MI}}$, at which 
the mixed phase transforms into one of the species-separated phases. 

\begin{figure*}[hb]
	\centering
	\includegraphics[width=0.75\linewidth]{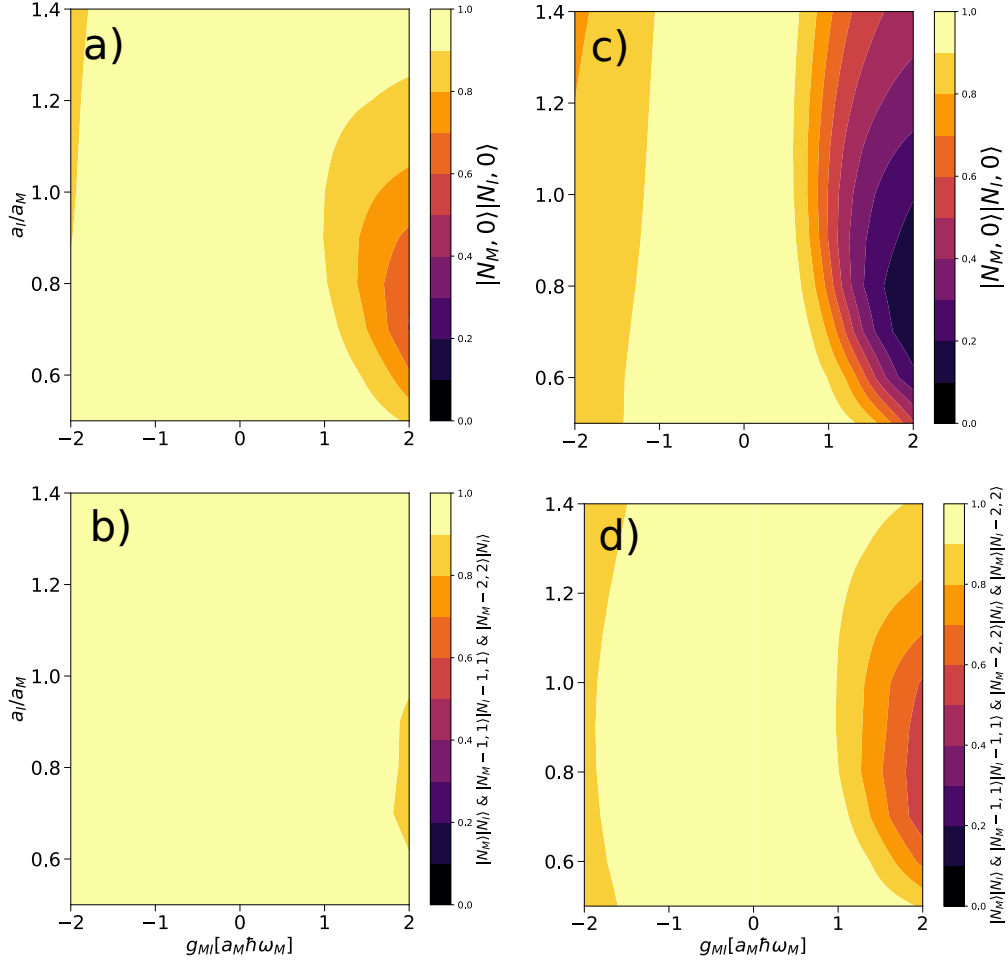}
	\caption{Projection amplitudes of the many-body ground state
	on number states 
	$|\bra{\vec{n}^{M}} \otimes \bra{\vec{n}^{I}}\ket{\Psi}|^2$
	for $N_B=5$ majority particles 
	and $N_I=1$ (first column) or $N_I=2$ (second column) impurities
	as a function of the inter-component coupling strength $g_{_{MI}}$
	and impurity localization $a_{I}/a_{M}=\sqrt{1/\eta}$ 
	with $\eta=\omega_{I}/\omega_{M}$ being trap frequency ratio 
	and $a_{\sigma}=\sqrt{\hbar/m \omega_{\sigma}}$ 
	the harmonic oscillator length of species $\sigma$.
	The SPFs constituting the permanents are eigenfunctions
	of the effective Hamiltonian \eqref{eq:projection_hamiltonian}.
	The first row corresponds to the projection 
	on the condensed number state 
	$\ket{N_B} \ket{N_I}$,
	while in the second row one sums over contributions 
	from two-particle excitations 
	$\ket{N_B-2,2} \ket{N_I}$, $\ket{N_B} \ket{N_I-2,2}$ 
	and $\ket{N_B-1,1} \ket{N_I-1,1}$.}
	\label{fig:number_states.eps}
\end{figure*}

In Fig.\ \ref{fig:number_states.eps}
we show the projection on number state $\ket{N_M}\ket{N_I}$ (first row)
and a sum over projections on the above mentioned permanents (second row)
for $N_B=5$ and $N_I=1$ (first column) or $N_I=2$ (second column).
For negative couplings the state $\ket{N_M}\ket{N_I}$ provides
a major contribution and the effective picture holds.
Let us focus in the following on positive couplings.
In Fig.\ \ref{fig:number_states.eps}a ($N_I=1$),
we observe that the state $\ket{N_M}\ket{N_I}$ 
has indeed a major contribution at coupling strength below $1.0$.
Once inter-species correlations build up with increasing coupling strength, 
the state $\ket{N_M-1,1}\ket{N_I-1,1}$ grows in importance,
which corresponds to a simultaneous 
single-particle excitation within each component.
This is mostly pronounced around $\eta_c$.
Double excitations within the majority species 
$\ket{N_M-2,2}\ket{N_I}$ are of minor amplitude
and rather of relevance for a localized impurity $a_I/a_M \ll 1$.
All in all, the low-lying excitations 
of the effective potentials \eqref{eq:projection_hamiltonian}
provide a good description (Fig.\ \ref{fig:number_states.eps}b).
In Fig.\ \ref{fig:number_states.eps}c ($N_I=2$), 
we observe that the state $\ket{N_M}\ket{N_I}$ 
loses its contribution very quickly
as one goes deeper into the regime of strong entanglement.
Although we are able to get a better understanding 
for weak entanglement by
including two-particle excitations mentioned above,
our effective picture clearly breaks down for strong entanglement.
There we may account only for as much as $\approx 50\%$ 
of the ground state, 
even though the one-body density in 
\eqref{eq:projection_hamiltonian} incorporates beyond
mean-field corrections.

\begin{figure*}[ht]
	\centering
	\includegraphics[width=0.8\linewidth]{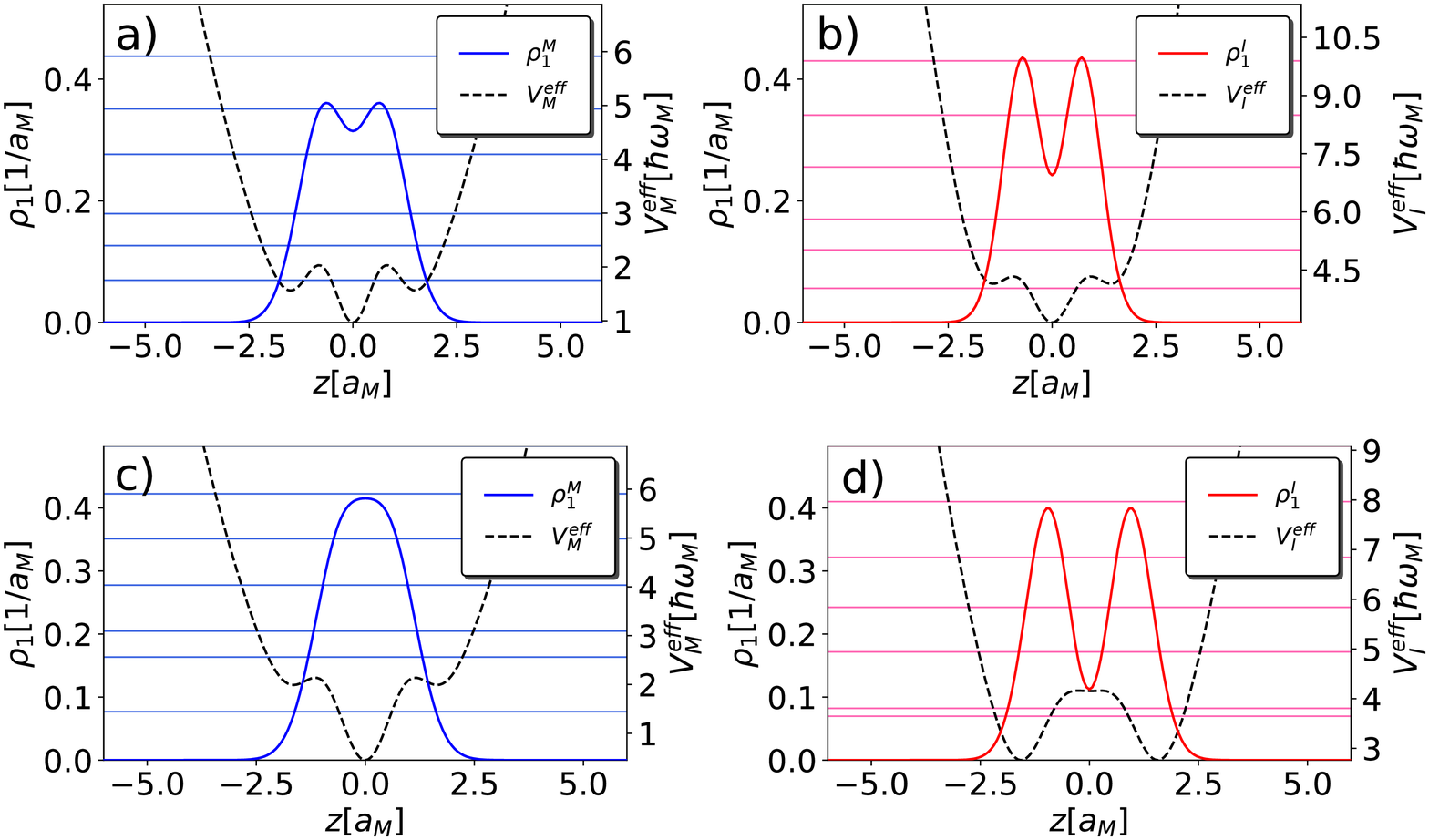}
	\caption{ML-X ground state densities $\rho_1^{\sigma}(z)$ inside
	effective one-body potentials $V_{\sigma}^{eff}(z)$ 
	from \eqref{eq:projection_hamiltonian} for
	$N_M=5$, $N_I=2$, $g_{_{MI}}=2$ and 
    $a_I/a_M=0.8$ (first row) or
    $a_I/a_M=0.9$ (second row).
	Horizontal lines are eigenenergies of \eqref{eq:projection_hamiltonian}.}
	\label{fig:eff_picture_breakdown_densities.eps}
\end{figure*}

Let us take a closer look at this regime, 
where the single-particle picture \eqref{eq:projection_hamiltonian} 
tends to break down.
We show in Fig.\ \ref{fig:eff_picture_breakdown_densities.eps}
the one-body densities for $N_B=5$ majority particles and
$N_I=2$ impurities in the strong entanglement region at $g_{_{MI}}=2$.
The first row corresponds to the $CF$ phase at $a_I/a_M=0.8$.
Here we recognize immediately, why the effective picture fails.
The origin of the two humps in the one-body density
is counter-intuitive considering that
they are at the position of local maxima of the effective potential.
One would rather expect a density profile 
with three peaks at the positions of the potential minima.

\begin{figure*}[ht]
	\centering
	\includegraphics[width=0.9\linewidth]{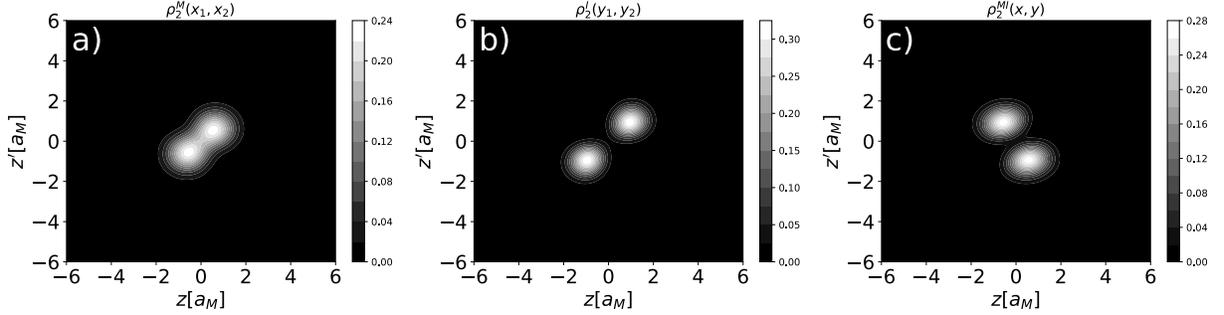}
	\caption{a) Majority two-body density matrix $\rho_2^{M}(x_1,x_2)$ 
	from eq.\ \eqref{eq:rho2}.
	b) Impurity two-body density matrix $\rho_2^{I}(y_1,y_2)$ 
	from eq.\ \eqref{eq:rho2}.
	c) Inter-species two-body density matrix $\rho_2^{MI}(x,y)$
	from eq.\ \eqref{eq:rho2inter}
	of the ground state for
	$N_M=5$, $N_I=2$, $g_{_{MI}}=2$ and $a_I/a_M=0.9$.}
	\label{fig:eff_picture_breakdown_dmats.eps}
\end{figure*}

The second row ($a_I/a_M=0.9$) seems at first glance
to be an $IMI$ phase. The majority is at the core, while
the impurity forms a shell.
Upon a more detailed investigation we notice
that the majority species is broader than it should be inside 
the squeezed "harmonic" trap.
The humps of the impurity also do not coincide with  the
positions of the minima of the respective effective potential.
As a matter of fact this phase is a latent $CF$ phase,
which becomes clear 
when we analyse the corresponding two-body density matrices 
in Fig.\ \ref{fig:eff_picture_breakdown_dmats.eps}.
The intra-species two-body density matrices \eqref{eq:rho2}
(Fig.\ \ref{fig:eff_picture_breakdown_dmats.eps}a-b) indicate
that particles of the same component 
avoid the trap center and form a cluster
either on the right or the left side w.r.t.\ trap center.
Moreover, the inter-species density matrix
\eqref{eq:rho2inter}
(Fig.\ \ref{fig:eff_picture_breakdown_dmats.eps}c)
tells us that the two different clusters of majority and impurity
will always be found on opposite sides of the trap
with a rather small spatial overlap between them. 
It allows to diminish the impact 
of the repulsive energy on the total energy 
at the cost of paying potential energy. 
These are clear signatures of the $CF$, 
which are blurred in the reduced one-body density.
We note that the parameter space where ML-X
predicts an $IMI$ phase, whereas $MF$ produces $SB$ phase,
we have in fact a latent $CF$, hidden behind a one-body quantity.

\begin{figure*}[ht]
	\centering
	\includegraphics[width=0.8\linewidth]{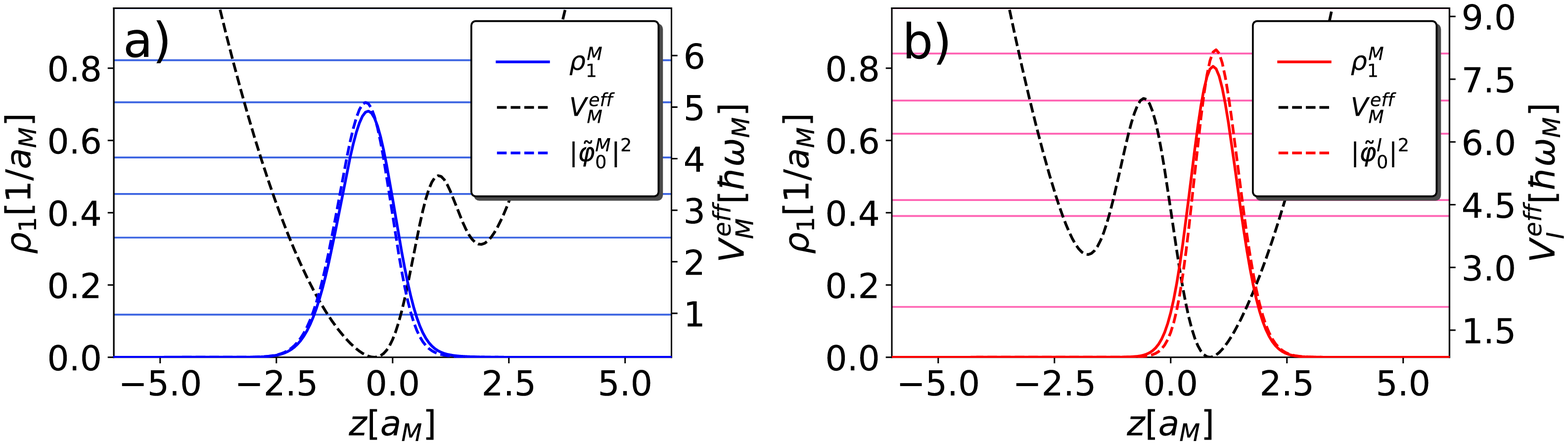}
	\caption{ML-X parity-broken ground state densities $\rho_1^{\sigma}(z)$ 
	obtained from \eqref{eq:hamiltonian} by slightly displacing 
	centers of harmonic traps in opposite directions
	for $N_M=5$, $N_I=2$, $g_{_{MI}}=2$ and 
    $a_I/a_M=0.9$.
	Induced one-body potentials $V_{\sigma}^{eff}(z)$ 
	are calculated from \eqref{eq:projection_hamiltonian}
	and $\tilde{\varphi}_0^{\sigma}(z)$ 
	are the corresponding ground states.
	Horizontal lines are eigenenergies 
	of \eqref{eq:projection_hamiltonian}.}
	\label{fig:SB.eps}
\end{figure*}

Above, we have mentioned that in the literature
the $CF$ phase was suggested to be a superposition
of two parity-broken mean-field states 
$\ket{\Psi} =c_1 \ket{N_M}_L \ket{0_M}_R \otimes \ket{0_I}_L \ket{N_I}_R
+ c_2 \ket{0_M}_L \ket{N_M}_R \otimes \ket{N_I}_L \ket{0_I}_R$
as a result of the degeneracy onset.
Indeed, ML-X has two prominent orbitals on the species layer
and two major SPFs on the particle layer. 
Nevertheless, the other occupied species orbitals and SPFs provide
a minor contribution, as we have evidenced in Fig.\ \ref{fig:entanglement.eps} second column,
where the entropies take values beyond $\ln(2)$.
To provide an illustrative example
we displace the trap centers in \eqref{eq:hamiltonian} by a small amount
to energetically separate the two symmetry broken configurations.
For parameter values for which the $CF$ phase is observed,
we perform again the improved relaxation to the find ground state of the system in order
to check whether it is indeed a MF state.
It turns out that
the majority species and the impurity species are still fragmented states
though the degree of depletion is much less
compared to the parity-symmetric ground state.
The species entropy $S_{vN}$ is greatly reduced, but still appreciable.
The impact of correlations is also visible
in Fig.\ \ref{fig:SB.eps}.
The ground state of the effective potential \eqref{eq:projection_hamiltonian} 
is different from the one-body density of the many-body ML-X wave-function. 
This is caused by
induced attractive interactions mediated by the inter-component coupling,
a beyond-mean field effect \cite{induced2018}.

To conclude our discussion about the high-entaglement regime, 
we state that the
mean-field approach, being an effective one-body model, 
fails to explain a one-body quantity such as reduced one-body density.
Nevertheless, it manages to characterize quite well
one of the two possible
configurations of the entangled many-body state.
The latter is not just a simple superposition 
of two mean-field states describing two different
parity-broken configurations. A thourough analysis showed
that on the many-body level the $SB$ phase
is in fact slightly entangled,
while each species is partially fragmented.
We also evidenced that $CF$ completely dominates
the highly correlated regime and made a link
of its appearance to the onset of $SB$ on the mean-field level.
Sometimes $CF$ is even camouflaged 
behind core-shell $IMI$ or $MIM$ densities,
indicating that the one-body density 
is not enough to distinguish between them.

\section{Phase separation: Impact of particle numbers}
\label{sec:many_body}

When increasing the number of majority atoms $N_M$, while keeping $N_I$ fixed, 
one might expect two properties based on an intuition 
for few-body systems.
First, the location of the strong entanglement regime  will be shifted
towards lower values of $a_I/a_M\approx\sqrt{1/\eta_c}=\sqrt[7]{N_I/N_M}$. 
Thus, the $IMI$ phase will cover the most part
of our parameter space for positive $g_{_{MI}}$.
Second, at a fixed $\eta$ the critical coupling $g_{_{MI}}^c$ for the miscible-immiscible transition
will decrease, because according to \eqref{eq:barrier_criterion_I} 
the majority species will be able 
to induce a barrier for the impurity species already for a much weaker coupling.
The induced barrier of the majority 
on the other hand will not be affected 
according to \eqref{eq:barrier_criterion_M}.

\begin{figure*}[ht]
	\centering
	\includegraphics[width=0.6\linewidth]{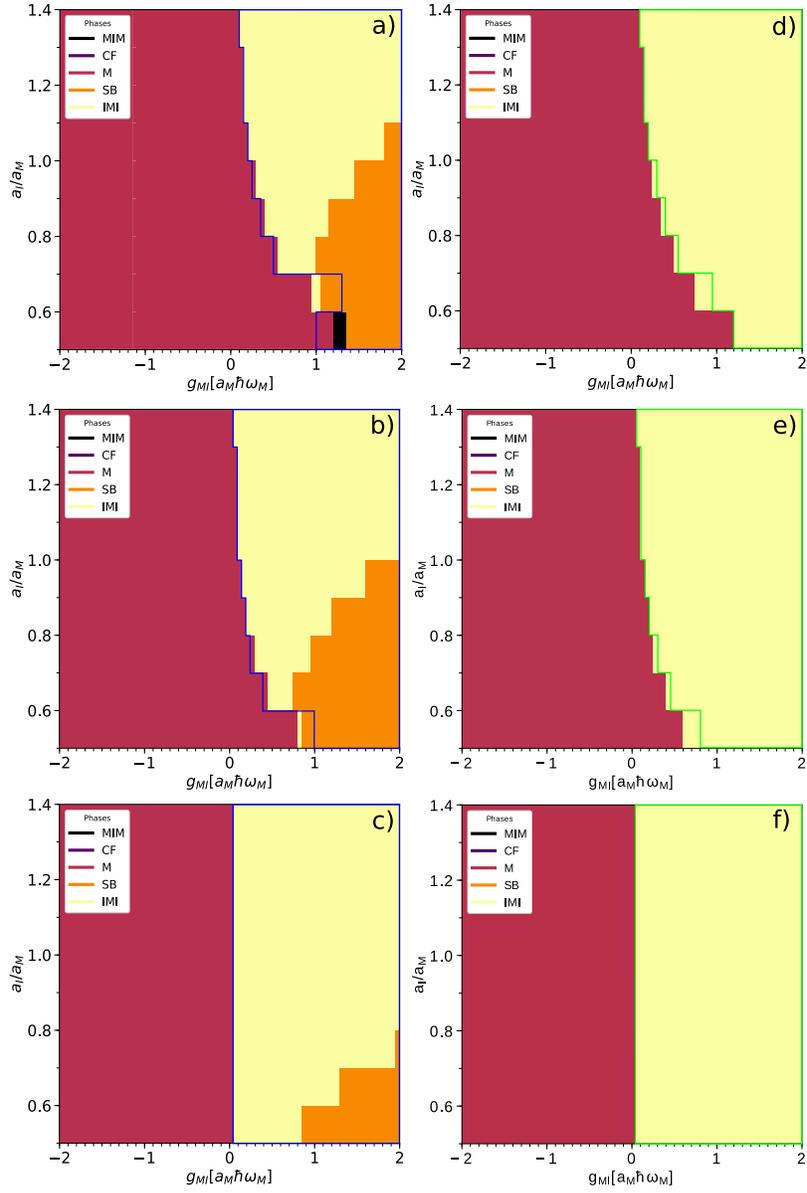}
	\caption{Mean-field (first column) and ML-X (second column) 
			ground state phase diagrams for $N_I=1$ impurity and
			$N_M=10$ (first row),
			$N_M=20$ (second row)
			or $N_M=1000$ (third row)
			as a function of the inter-component
			 coupling strength $g_{_{MI}}$
			and impurity localization $a_{I}/a_{M}=\sqrt{1/\eta}$ 
			with $\eta=\omega_{I}/\omega_{M}$ being the trap frequency ratio 
			and $a_{\sigma}=\sqrt{\hbar/m \omega_{\sigma}}$ 
			the harmonic oscillator length of species $\sigma$. 
			The nomenclature of the phases is as follows: 
			$M$ for miscible, 
			$MIM$ for core-shell with impurity at the core, 
			$IMI$ for core-shell with majority at the core,
			$CF$ for composite fermionization and
			$SB$ for a phase with broken parity symmetry.
			The blue solid curve (first column) represents 
			the miscible-immiscible phase boundary 
			according to \eqref{eq:critical_g}.
			The green solid curve (second column) 
			is the miscible-immiscible phase boundary 
			based on mean-field.
			The coarse structure is due to the finite step-size of our data w.r.t.\ $a_I/a_M$.}
	\label{fig:many_body.eps}
\end{figure*}

Indeed, this is what we observe in the phase diagrams 
depicted in Fig. \ref{fig:many_body.eps}.
In the mean-field (first column) 
the location of the $SB$ phase relocates from $\sqrt{1/\eta_c}\approx0.79$ 
($N_M=5$ Fig.\ \ref{fig:MF_few_body.eps}a) to
$\sqrt{1/\eta_c}\approx0.72$ ($N_M=10$ Fig. \ref{fig:many_body.eps}a), 
then to $\sqrt{1/\eta_c}\approx0.65$ ($N_M=20$ Fig. \ref{fig:many_body.eps}b) 
and finally moves outside our parameter space 
$\sqrt{1/\eta_c}\approx0.37$ ($N_M=1000$ Fig. \ref{fig:many_body.eps}c). 
The blue curve, 
which estimates the miscible-immiscible transition according to 
\eqref{eq:critical_g} 
is in good agreement (except for the critical region $\eta_c$) 
with the mean field phase boundary.
We also recognize that for a fixed trap ratio $\eta$, 
the critical coupling strength
$g_{_{MI}}^c$ decreases with increasing $N_M$ and at $N_M=1000$
a very small $g_{_{MI}}^c<0.05$ is sufficient to cause phase separation,
which is below our resolution.

We have also performed the corresponding ML-X calculations (second column)
with $S=s_{\sigma}=6$ (first row), $S=s_{\sigma}=4$ (second row) and
$S=s_{\sigma}=2$ (third row) orbitals.
We remark that the latter case might not be converged to the exact solution, 
which is beyond numerical capabilities to verify. 
Still it provides valuable beyond mean-field corrections.
The deviations to the mean-field, still clearly visible at $N_M=10$, are
most pronounced near $\eta_c$.
They become less as the particle-imbalance is increased until
finally at $N_M=1000$ the phase diagrams almost coincide except
for a small $SB$ region. 
This is mainly attributed to the fact that the strong entanglement regime,
where deviations are to be expected, moves outside our parameter space
($a_I/a_M<0.5$).
Furthermore, the deviations may still be there, 
but on a finer coupling scale $g_{_{MI}}<0.05$ according to
\eqref{eq:barrier_criterion_M} and \eqref{eq:barrier_criterion_I}.

\section{Conclusions} 
\label{sec:conclusions}

In this work we have investigated the phase-separation
of a quasi-1D inhomogeneous Bose-Bose mixture 
in a three dimensional parameter space spanned by the
inter-component coupling $g_{_{MI}}$, 
harmonic length scale ratio $a_I/a_M=\sqrt{1/\eta}$
and the particle number ratio $N_I/N_M$, 
when the intra-component couplings $g_{\sigma}$ are switched off.
Although we have concentrated on the case of equal masses, 
our results may be easily extended to the more general case 
of unequal masses.
We expect some quantitative changes,
but the qualitative picture 
and the line of argumentation will remain unchanged.

The commonly used separation criterion $g_{_{MI}}>\sqrt{g_{_M} g_{_I}}$,
which is valid for homogeneous mixtures,
would predict a miscible-immiscible transition
for any finite coupling $g_{_{MI}}>0$.
However, this separation rule does not apply here, 
since we have harmonic traps
of different length scales.
We have analyzed the mechanism, 
which leads to phase separation, by using an effective mean-field picture.
Within this description each species is subject 
to an additional induced potential caused by the other component.
This potential has initially a Gaussian shape 
and grows linearly with the coupling strength $g_{_{MI}}$.
However, it does not immediately trigger a barrier
at the center of the harmonic trap. 
In fact, the species, which first manages to induce a barrier
for the other component upon increasing the coupling $g_{_{MI}}$, 
will stay at the center of its parabolic trap.
Meanwhile the other species will split up,
once the ground state energy 
of the effective potential drops below the barrier height. 
Thus, we end up with either a core-shell $IMI$ or a core-shell $MIM$ phase,
except for a highly competitive region, 
where the barrier conditions can be met simultaneously for both components.
We have derived a simple rule to predict the type of phase separation,
developed a straightforward algorithm 
to identify the miscible-immiscible phase boundary $g_{_{MI}}^c$
and gave a rough estimate 
on the phase boundary between the segregated phases 
$g_{_{MI}}^{^{SB}}$.

As a next step, we compared mean-field (MF) results 
to the numerically exact many-body calculations based on
Multi-Layer Multi-Configurational 
Time-Dependent Hartree Method for atomic mixtures (ML-X).
It turns out that MF agrees well with ML-X far away from 
the critical region $\sqrt{1/\eta_c}=\sqrt[7]{N_I/N_M}$.
At $\eta_c$ there are considerable quantitative deviations
and sometimes the two methods do not even agree 
on the type of phase separation.
This is caused by the growing inter-particle correlations,
which generate entanglement between the components
and increase the degree of species fragmentation.
We have seen that symmetry-broken phase ($SB$) 
is replaced by composite fermionization ($CF$), 
which is an entangled parity symmetric ground state. 
Furthermore, we have linked the onset of $SB$ 
to the fact that the entanglement entropy reaches a certain threshold
and saw a clear breakdown of the effective single-particle picture
in the strong entanglement region
in terms of a corresponding number state analysis.
This led to the discovery of a latent $CF$ phase in the $IMI$ region.
The latent $CF$ phase has the characteristic one-body density 
of the $IMI$ phase,
but a thorough analysis of the two-body densities 
reveals typical $CF$ features.
We have argued that at a finite coupling $g_{_{MI}}$ the $CF$
is not a simple superposition of two $SB$ states given by mean-field.

We have studied the impact of particle number
variations, which confirmed our intuition
that $\eta_c$ and thus the location of the strong entanglement regime
can be manipulated as a function of the particle number ratio.
Furthermore, for a fixed particle number ratio 
the critical coupling $g_{_{MI}}^c$ 
of the miscible-immiscible transition can be tuned
to lower values by increasing the number of particles 
while keeping the particle number ratio fixed.

Finally, we remark that an intriguing next step would be
to perform a similar study
of phase-separation at finite intra-component coupling $g_{\sigma}$.
The broadening or shrinking of the density profiles, 
depending on the sign and strength of $g_{\sigma}$,
will definitely modify the barrier conditions
\eqref{eq:barrier_criterion_M} and \eqref{eq:barrier_criterion_I}.
Another interesting but challenging direction 
would be the non-equilibrium dynamics by quenching the trap ratio
across the phase boundaries.

\begin{acknowledgments}
M.\ P.  acknowledges fruitful discussions with K.\ Keiler and M.\ Roentgen. 
M.\ P.  gratefully  acknowledges  a  scholarship  
of  the  Studienstiftung des deutschen Volkes.
\end{acknowledgments}

\setcounter{equation}{0}
\setcounter{figure}{0}
\setcounter{table}{0}
\renewcommand{\theequation}{S\arabic{equation}}
\renewcommand{\thefigure}{S\arabic{figure}}
\renewcommand{\thetable}{S\arabic{table}}

\appendix


%
\bibliographystyle{apsrev4-1}

\end{document}